\documentclass[12pt]{article}
\usepackage{amsmath}
\usepackage[psamsfonts]{amssymb}
\usepackage[dvips]{graphicx}
\usepackage[hypertex]{hyperref}
\numberwithin{equation}{section}
\newcommand{\la}{\lambda}

\newcommand{\Ncal}{\mathcal{N}}
\newcommand{\del}{\partial}
\DeclareMathOperator*{\Tr}{{\rm Tr}}

\DeclareMathOperator{\arccosh}{{\rm arccosh}}

\newcommand{\vp}{\varphi}
\newcommand{\Fcal}{{\cal F}}

\newcommand{\Hcal}{{\cal H}}
\newcommand{\thet}{\theta_k}
\renewcommand{\th}{\Theta}
\begin{document}
%%% Title page %%%%%
\begin{titlepage}

 \renewcommand{\thefootnote}{\fnsymbol{footnote}}
\begin{flushright}
 \begin{tabular}{l}
 {\tt hep-th/0610275}\\
 UT-06-22 \\
 IHES/P/06/53 \\
 \end{tabular}
\end{flushright}

 \vfill
 \begin{center}
 \font\titlerm=cmr10 scaled\magstep4
 \font\titlei=cmmi10 scaled\magstep4
 \font\titleis=cmmi7 scaled\magstep4
{\titlerm Correlator of Fundamental and Anti-symmetric \\
 Wilson Loops in AdS/CFT Correspondence}
 \vskip 1.9 truecm

\noindent{{\large
  Ta-Sheng Tai$^a$\footnote{E-mail:tasheng@hep-th.phys.s.u-tokyo.ac.jp}
and Satoshi Yamaguchi$^b$\footnote{E-mail:yamaguch@ihes.fr} }}
\bigskip
 \vskip .9 truecm
\centerline{\it $^a$ Department of Physics, Faculty of Science, 
University of Tokyo}
\centerline{\it Hongo, Bunkyo-ku, Tokyo 113-0033, JAPAN}
\bigskip
 \centerline{\it $^b$ IHES, Le Bois-Marie, 35, Route de Chartres
F-91440 Bures-sur-Yvette, FRANCE}
\vskip .4 truecm
\end{center}
 \vfill
\vskip 0.5 truecm

\begin{abstract}
We study the two circular Wilson loop correlator in which one is of 
anti-symmetric representation, while the other is of fundamental 
representation in 4-dimensional ${\cal N}=4$ super Yang-Mills theory. 
This correlator has a good AdS dual, which is a system of a D5-brane and 
a fundamental string. We calculated the on-shell action of the string, 
and clarified the Gross-Ooguri transition in this correlator. 
Some limiting cases are also examined. 
\end{abstract}

\vfill
\vskip 0.5 truecm

\setcounter{footnote}{0}
\renewcommand{\thefootnote}{\arabic{footnote}}
\end{titlepage}

\newpage

\tableofcontents
%%%%%%%%%%%%%%%%%%%%%%%%%%%%%%%%%%%%%%%%%%%%%%%%%%%%%%%%%%%%%%%%%%%%%%
\section{Introduction and summary}

The Wilson loop is one of the most interesting quantities which appeared in 
the AdS/CFT correspondence 
\cite{Maldacena:1997re}. 
In 
the computation of the expectation value of the Wilson loop, the stringy
effect (not just the supergravity) is essential. Actually, it is
calculated on the AdS side as the on-shell action of a classical
macroscopic string\cite{Rey:1998ik,Maldacena:1998im} in the
limit of large $N$ and large 't Hooft coupling $\lambda=4\pi g_s N=g^2_{YM} N$.

Recently, certain D-brane pictures of Wilson loops 
have drawn wide attention because 
they are powerful in evaluating the Wilson loop of higher rank 
representation. The D3-brane with electric flux, which has already been 
investigated in \cite{Rey:1998ik}, is found to 
describe Wilson loops of symmetric representations (or multiply wound 
Wilson loops) by Drukker and Fiol \cite{Drukker:2005kx}. On the 
other hand, the Wilson loop of anti-symmetric representation is 
found to be described by a D5-brane with electric 
flux\cite{Hartnoll:2006hr,Yamaguchi:2006tq,Gomis:2006sb}. 
In both 
cases, the results show successful agreement with the gauge theory side 
calculations \cite{Erickson:2000af,Drukker:2000rr}. 
Other further investigations along this line can be found in \cite{Rodriguez-Gomez:2006zz,Okuyama:2006jc,Hartnoll:2006is,
Hartnoll:2006ib,Chen:2006iu,Giombi:2006de}.

One of the interesting phenomena in the AdS picture of Wilson loops is the 
phase transition, which occurs in the two Wilson loop correlator. 
This is first anticipated by Gross and Ooguri \cite{Gross:1998gk}, and 
further explored in 
\cite{Zarembo:1999bu,Olesen:2000ji,Kim:2001td,Arutyunov:2001hs,Drukker:2005cu,Tsuji:2006zn,Ahn:2006px}. 
If the two loops are close enough, the classical string of annulus
topology leads to the smallest action, which dominates the connected 
correlator. By contrast, when the two loops get far enough, 
two disk solution with massless propagating modes dominates the connected 
correlator. Between them, there exists a critical point. 

In this paper, we study two concentric circular Wilson loop correlator 
of different representations, i.e. one is fundamental and the other is 
anti-symmetric%
\footnote{Anti-symmetric--anti-symmetric correlator 
has been studied in \cite{Ahn:2006px}.}. Because the circular 
anti-symmetric Wilson loop has a gravity dual as a D5-brane with 
electric flux, this correlator can thus be recognized as a system of a 
D5-brane and an F-string. More explicitly, it is realized as a 
string which propagates between the AdS boundary and the 
D5-brane. 

Here let us summarize the results of this paper. We calculated the correlator
by evaluating the on-shell action of the string worldsheet. This correlator also
exibits the phase transition. The phase structure is shown in figures~\ref{pi/2}, \ref{2pi/3} and \ref{pi/3}. The critical surface of the phase transition is represented by a solid line. There is also a critical surface represented by a dashed line where the annulus solution becomes unstable.

The parameter $\th$ (direction in $S^5$) dependence is quite nontrivial and interesting. As easily seen from the figures \ref{pi/2}, \ref{2pi/3} and \ref{pi/3}, 
as well as from the expression in section 3, the correlator has
the symmetry $\th\to 2\theta_k-\th$, where $\theta_k$ is a function of $k$ 
as in eq.\eqref{thetak}. Especially, the point $\th=2\theta_k-\pi$ 
looks similar to the point $\th=\pi$. At $\th=\pi$, it is always in the disk 
phase, since it is BPS even when two loops get on top of each other. 
On the other hand, $\th=2\theta_k-\pi$ is not a priori related to the BPS configuration, but it is
always in the disk phase. This fact implies that overlapped loops at 
$\th=2\theta_k-\pi$ bear a similar nature to the BPS loops. 

Another interesting feature related to $\th$ dependence is the maximum 
of the critical distance. The critical distance is maximal at $\th=\theta_k$. 
This is reasonable in the AdS picture 
since the D5-brane sits at $\theta=\theta_k$. However, 
it is rather misterious because 
the origin of $\theta_k$ is quite different from $\th$ on the gauge theory side. 
It should be interesting and useful to consider
the origin of $\theta_k$ more geometrically on the gause theory side.

We also studied a couple of limiting cases. The solution 
of two fundamental Wilson loop correlator first examined by 
Zarembo\cite{Zarembo:1999bu} is recovered, meanwhile the Coulomb-like behavior 
is found 
when the separation is much smaller than the circle radii. 

As a future work, the Yang-Mills side analysis like 
\cite{Zarembo:2001jp,Plefka:2001bu} seems to be interesting. 
Correlators of Wilson loops of other representations by using the 
D3-brane \cite{Drukker:2005kx} or supergravity solutions 
\cite{Yamaguchi:2006te,Lunin:2006xr} are as well worthy of study. 
Also, the extension to finite temperature cases is in progress. 

The rest of this paper is organized as follows. In sec.2, the detailed 
setup is given. In sec.3, the AdS side calculation is performed 
through a series of elliptic integrals. Finally, in sec.4, 
we plotted the phase diagram, which visualizes clearly the Gross-Ooguri 
transition. 
Comments on these solutions are also provided, where known 
results appear as limiting cases. Some definitions and algebra 
involving elliptic integrals are added in the appendix \ref{app-a}. 
The explicit forms of the Coulomb coefficients are shown in appendix 
\ref{app-coulomb}.

\section{Setup}
In $\Ncal=4$ Super Yang-Mills theory, the Wilson loop is defined as
\begin{align}
W_{\mathsf{R}}(C):= \Tr_{\mathsf{R}}\big[
P \exp \oint_{C} d\tau (i A_{\mu}{\dot x}^{\mu}+\Phi_{I}n^{I}(\tau)|\dot x|)
\big],
\end{align}
where $A_{\mu}$ is the gauge field, $\Phi_I$'s are six scalar fields, 
$C$ is a closed trajectory in 4-dimensional spacetime parameterized by 
$\tau$, $n_I$ is a 6-dimensional unit vector, and $\mathsf{R}$ is a representation of SU($N$).  In this paper, we
consider the following type of correlation function
\begin{align}
 \left\langle W_{A_k}(C_1)W_{\Box}(C_2) \right\rangle. \label{cor}
\end{align}
We take the trajectories $C_1$ and $C_2$ as two circles. Both of them 
are embedded in a 3-dimensional hyperplane of 4-dimensional space. They 
are parallel and the centers of them are on the axis which is 
orthogonal to the circles. If we introduce the Cartesian coordinate 
$x_{\mu}$, $C_1$ is expressed via $\tau$ as 
\begin{align}
 x_{\mu}=(0,R_1 \cos \tau,R_1 \sin \tau,0),\qquad
n^{I}=(1,0,0,0,0,0),
\end{align}
while $C_2$ is 
\begin{align}
 x_{\mu}=(L ,R_2 \cos \tau,-R_2 \sin \tau,0),\qquad
 n^{I}=(\cos \th,\sin \th,0,0,0,0),\qquad (0<L, ~ 0\le \th \le \pi).
\label{c1}
\end{align}
So parameters of this configuration are $R_1,R_2,L,\th$.

The correlator \eqref{cor} factorizes in the leading term of 
the $1/N$ expansion. But this term does not depend on parameters 
$L,R_1,R_2,\th$. In order to see their dependence, 
the subleading terms become relevant. In other words, what we have to consider 
is the ``connected correlator''
\begin{align}
\left\langle W_{A_k}(C_1)W_{\Box}(C_2) \right\rangle_{conn}
=\left\langle W_{A_k}(C_1)W_{\Box}(C_2) \right\rangle-
\left\langle W_{A_k}(C_1) \right\rangle
 \left\langle W_{\Box}(C_2) \right\rangle. 
\end{align}
As we will see later, it is convenient to define the function 
$V_{k}(R_1,R_2,L ,\th)$ as
\begin{align}
 V_{k}(R_1,R_2,L ,\th):=
\frac{\left\langle W_{A_k}(C_1)W_{\Box}(C_2) \right\rangle-
\left\langle W_{A_k}(C_1) \right\rangle
 \left\langle W_{\Box}(C_2) \right\rangle}{
\left\langle W_{A_k}(C_1) \right\rangle}.\label{Vk}
\end{align}
On the AdS side, this correlation function is described by a system of
 a D5-brane and a fundamental string subject to certain boundary conditions.
 The D5-brane is rigid because it is much heavier than the fundamental
 string. By contrast, the string is flexible, 
so various kinds of contributions should be considered.

\begin{figure}
 \begin{center}
  \includegraphics[width=7truecm]{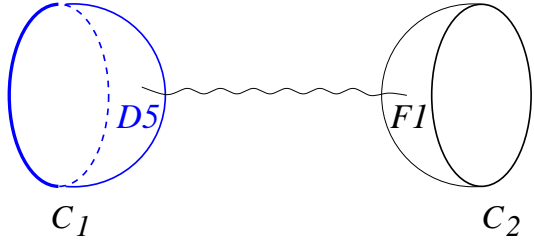}
 \end{center}
\caption{The schematical picture of the disk solution with massless propagating 
modes. 
The disk solution is of order $1/g_s$, the massless propagator 
is of order $g_s^2$ and the coupling to the D5-brane is of order 
$1/g_s$. Totally, the contribution is of order $g_s^0$, which is the same
order as the annulus contribution.}
\label{fig-disk}
\end{figure}

\begin{figure}
 \begin{center}
  \includegraphics[width=7truecm]{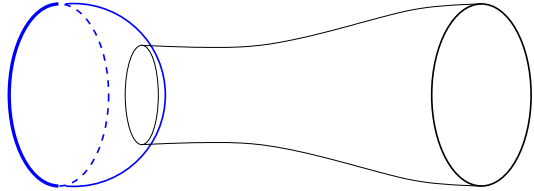}
 \end{center}
\caption{The schematical picture of the annulus. Since the D5-brane is much 
heavier than the fundamental string, the D5-brane is not deformed. The 
contribution from this annulus solution is as well of order $g_s^0$.}  
\label{fig-annulus}
\end{figure}

There are two possible leading contributions to $V_k$. 
One is a disk bounded by $C_2$ 
with massless propagating modes connecting the F-string 
and the D5-brane (see figure \ref{fig-disk}). 
The other is 
an annulus whose boundaries are 
attached to the D5-brane and $C_2$ (see figure \ref{fig-annulus}). 
These two contributions are of 
the same order $\simeq{\cal O}(N^0)$ in $1/N$ expansion. 
Their magnitudes are estimated by the on-shell action of 
the annulus and the disk%
\footnote{The massless propagator contributes a factor of order ${\cal O}(\lambda^n)$ 
to the amplitude. While is compared to $\exp(-\sqrt{\lambda})$, it is thus 
negligible in large $\lambda$ limit.}, respectively, in large $\lambda$ limit.
In summary, in large $\lambda$ limit, $V_k$ is approximated as
\begin{align}
 V_{k}(R_1,R_2,L ,\th)=\exp(-S_{min}),
 \label{logV}
\end{align}
where $S_{min}$ is the global minimum of the string action. 
The classical on-shell action of the disk, which corresponds to a single 
circular Wilson loop, has been calculated in \cite{Berenstein:1998ij}. The 
result is 
\begin{align}
 S_{disk,min}=-\sqrt{\lambda}. 
\label{sdisk}
\end{align}
The question now is the competition of 
\eqref{sdisk} and that of the annulus. We will 
investigate the classical annulus solution and 
compare it with \eqref{sdisk} in this paper.

Before doing this, 
we can make some observation upon using the conformal symmetry. 
There are certain constraints on the correlation function. 
For example, we have a relation from the dilatation 
\begin{align}
 V_{k}(R_1,R_2,L ,\th)=V_{k}(a R_1,a R_2,a L ,\th),
\end{align}
where $a$ is a real positive parameter of the dilatation. We can also use 
the special conformal transformation with a vector parameter $b^{\mu}$ 
\begin{align}
 x'^{\mu}=\frac{x^{\mu}-b^{\mu} |x|^2}{1-2b^{\mu}x_{\mu}+|b|^2|x|^2}.
 \label{sc}
\end{align}
This special conformal transformation \eqref{sc} along $x_1$, i.e. 
$b^{\mu}=(h,0,0,0)$ makes 
\begin{align}
\begin{aligned}
& V_{k}(R_1,R_2,L ,\th)=V_{k}(R_1',R_2',L ',\th),\\
& R_1'=\frac{R_1}{1+h^2 R_1^2},\qquad 
R_2'=\frac{R_2}{1-2hL +h^2(L ^2+R_2^2)},\\
&L '=\frac{L -h(L ^2+R_2^2)}{1-2hL +h^2(L ^2+R_2^2)}
 +\frac{hR_1^2}{1+h^2 R_1^2}.
\end{aligned}
\end{align}
Therefore, $V_{k}$ is a function of the invariant combination 
$\eta$ as%
\footnote{This $\eta$ plays the same role as the ``cross ratio.''}
\begin{align}
 V_{k}(R_1,R_2,L ,\th)=\widetilde V_{k}(\eta,\th),\qquad
\eta=\frac{R_1^2+R_2^2+L ^2}{2R_1R_2}.
\label{eta}
\end{align}
On the AdS side, 
this simple dependence on $(\eta,\th)$ 
is not quite trivial. This is due to the introduction of the cutoff, which makes 
the conformal symmetry not manifest. In order to check this conformal symmetry,
we keep the parameters $(R_1,R_2,L)$ generic in the calculation in the next section.
We will find that the constraint \eqref{eta} is actually satisfied.

\section{Classical solution of the annulus}\label{sec-solve}
In this section, we will investigate the annulus solution. 
We use similar techniques employed by \cite{Zarembo:1999bu,Olesen:2000ji,
Kim:2001td}. 
We write down the solution and the corresponding on-shell action. 
One can see the ``phase structure'' from the result. 

\subsection{Ansatz and equations of motion}
In this paper, we use the following notation for 
$AdS_5\times S^5$ with radius $\sqrt{\alpha'}\lambda^{1/4}$ 
\begin{align}
 ds^2=\frac{\alpha'\sqrt{\lambda}}{y^2}(dy^2+dr^2+r^2d\vp^2+dt^2+dx^2)
+\alpha'\sqrt{\lambda}(d\theta^2+\sin^2\theta d\Omega_4^2). \label{metric}
\end{align}
The boundary of $AdS_5$ is at $y=0$. The coordinate 
$(r,\vp,t,x)$ 
at the boundary 
is related to the Cartesian coordinate $x_{\mu}$ 
introduced in the previous section as 
\begin{align}
 x_1=x, \qquad  x_2=r \cos \vp,\qquad x_3=r\sin \vp,\qquad x_4=t.
\end{align}
The string dual of $W_{A_k}(C_1)$ is a D5-brane
 \cite{Yamaguchi:2006tq,Gomis:2006sb}. 
Its explicit form\cite{Pawelczyk:2000hy,Camino:2001at} is expressed as
\begin{align}
  r^2+y^2=R_1^2,\qquad \theta=\theta_k, \label{D5r}
\end{align}
where $R_1$ is the radius of $C_1$, and $\theta_k$ is related to $k$ as
\begin{align}
 \frac{k}{N}=\frac{1}{\pi}\left(\theta_k-\frac12 \sin 2\theta_k\right).
\label{thetak}
\end{align}
The gauge field excitation on the D5-brane worldvolume has the 
field strength 
\begin{align}
\Fcal_{y\vp}=-i\cos\thet \frac{\alpha'\sqrt{\lambda} R_1}{y^2},
\qquad \Fcal:=2\pi \alpha' dA.
\label{D5A}
\end{align}
The worldsheet stretching between this D5-brane and $C_2$ is 
symmetric under the rotation $\vp\to\vp+\alpha,\ 
\alpha$: real constant parameter. Hence, 
we can assign this symmetry to our solution. 
The reparameterization degrees of freedom enable one to set 
the worldsheet coordinates as $(\sigma,\vp)$, 
where $\vp$ is the same as the spacetime $\vp$. 
The ansatz used is written as
\begin{align}
y&=y(\sigma),\qquad r=r(\sigma),\qquad x=x(\sigma),\qquad \theta=\theta(\sigma).\label{ansatz}
\end{align}
In this ansatz, we include $\sigma$ dependence of $\theta$ 
because the two boundaries are separated in $S^5$ in general, i.e. 
the D5-brane is wrapped on an $S^4$ at $\theta=\theta_k$, 
while $C_2$ sits at a point $\theta=\th$ in $S^5$. This kind of separation in $S^5$
is first considered extensively in \cite{Drukker:2005cu}. 
Their method of solving the equation of motion is applicable to the problem here, 
if one uses the conformal symmetry and simplifies the configuration.  Here, however, we do not use the conformal symmetry to simplify the configuration. This is because one purpose of this calculation is to check the conformal symmetry.

The bulk part comes from the Nambu-Goto action
\begin{align}
 S_{bulk}=\frac{1}{2\pi \alpha'}\int d \sigma d\vp \sqrt{\det G},
\end{align}
where $G$ is the induced metric. Plugging \eqref{ansatz} 
and integrating out $\vp$, we obtain 
\begin{align}
 &S_{bulk}=\sqrt{\la} \int d \sigma \frac{r}{y^2}
\sqrt{x'^2+r'^2+y'^2+y^2\theta'^2} ,\label{bulk-action}
\end{align}
where prime ${}'$ denotes the $\sigma$ derivative.

At both boundaries, we need to include 
proper boundary terms. Let $\sigma=\sigma_1$ be the boundary on the D5-brane, 
whereas $\sigma=\sigma_2$ be the boundary at $C_2$ with $\sigma_1 < \sigma_2$. 
At $\sigma=\sigma_1$, the boundary is constrained on the rigid D5-brane in \eqref{D5r},
and should satisfy at least the following conditions
\begin{align}
 r^2+y^2=R_1^2,\qquad x=0,\qquad \theta=\theta_k.
\label{bc1-1}
\end{align}
These are not all the conditions however. 
Due to the gauge field excitation in \eqref{D5A},
 we should add 
\begin{align}
\begin{aligned}
 S_{bdy,1}=i\oint_{\sigma=\sigma_1} 
  d\varphi ~A_{\varphi} ~ +(\text{constant})=\sqrt{\lambda} 
  \cos{\theta_k} \left(1-\frac{R_1}{y}\right),~
    &&A_{\varphi}=\frac{1}{2\pi \alpha'}i  \cos{\theta_k}
   \frac{\alpha'\sqrt{\lambda} R_1}{y}.
 \label{gaugevol}
 \end{aligned}
 \end{align}
The (constant) ensures that $S_{bdy,1}=0$ when $y=R_1$, where 
the circle of $\vp$ shrinks to a point. 
The boundary condition is derived from the variational principle.
The variation of the bulk action has boundary terms at $\sigma=\sigma_1$ as 
\begin{align}
 \delta S_{bulk}|_{bdy,1}
=&-p_r \delta r-p_y \delta y-p_x \delta x-p_{\theta}\delta \theta.
\label{bv}
\end{align}
The momenta $p$ are written as
\begin{align}
p_r=\sqrt{\lambda}\frac{r}{y^2}\frac{r'}{\sqrt{x'^2+r'^2+y'^2+y^2\theta'^2}},\\
p_y=\sqrt{\lambda}\frac{r}{y^2}\frac{y'}{\sqrt{x'^2+r'^2+y'^2+y^2\theta'^2}},\\
p_x=\sqrt{\lambda}\frac{r}{y^2}\frac{x'}{\sqrt{x'^2+r'^2+y'^2+y^2\theta'^2}},\\
p_{\theta}=\sqrt{\lambda}\frac{r\theta'}{\sqrt{x'^2+r'^2+y'^2+y^2\theta'^2}}.
\end{align}
Due to $r^2+y^2=R_1$, 
it is convenient to rewrite \eqref{bv} as 
\begin{align}
 \delta S_{bulk}|_{bdy,1} &=-\frac12 \left(\frac{p_r}{r}+\frac{p_y}{y}\right)
\left(r\delta r+y\delta y\right)
-\frac12 \left(\frac{p_r}{r}-\frac{p_y}{y}\right)
\left(r\delta r-y\delta y\right)
-p_x \delta x-p_{\theta}\delta \theta.
\end{align}
On the other hand, the variation of $S_{bdy,1}$ is written as (using the
condition $r\delta r+y\delta y=0$)
\begin{align}
 \delta S_{bdy,1}=-\frac12 \sqrt{\lambda}\cos \theta_k \frac{R_1}{y^3}
(r\delta r-y\delta y).
\end{align}
We obtain totally 
\begin{align}
 \delta S_{bulk}|_{bdy,1}+\delta S_{bdy,1}
 &=-\frac12 \left(\frac{p_r}{r}+\frac{p_y}{y}\right)
\left(r\delta r+y\delta y\right)
-\frac12 \left(\frac{p_r}{r}-\frac{p_y}{y}
+\sqrt{\lambda} \cos \theta_k \frac{R_1}{y^3}\right)
\left(r\delta r-y\delta y\right)\nonumber\\
&\qquad-p_x \delta x-p_{\theta}\delta \theta.
\end{align}
The first, third and fourth term vanish because of 
\eqref{bc1-1}. From the second term, it can be seen that 
the additional boundary condition at $\sigma=\sigma_1$ is 
\begin{align}
 \frac{p_r}{r}-\frac{p_y}{y}+\sqrt{\lambda} \cos \theta_k \frac{R_1}{y^3}=0.
 \label{tb}
\end{align}

Next, let us turn to the other boundary at $\sigma=\sigma_2$. Naively, the area of
the minimal surface is infinite. It is necessary to 
introduce a cutoff at 
$y=\epsilon$ in order to regularize the action. 
We follow \cite{Drukker:1999zq} and perform the
 Legendre transformation. This gives a boundary term
\begin{align}
 S_{bdy,2}=-p_y y-p_{\theta}(\theta-\th), ~~~~\text{at}~\sigma=\sigma_2.
\label{bt2}
\end{align}

Now we solve the equations of motion subject to the above boundary
conditions.  Each equation of motion of $r,y,x,\theta$, respectively,
derived from \eqref{bulk-action} can be listed as
\begin{align}
& \frac{1}{y^2}\sqrt{x'^2+r'^2+y'^2+y^2\theta'^2}-\del_{\sigma}
\left(\frac{r}{y^2}\frac{r'}{\sqrt{x'^2+r'^2+y'^2+y^2\theta'^2}}\right)=0,\label{eom-r}\\
&-2\frac{r}{y^3}\sqrt{x'^2+r'^2+y'^2+y^2\theta'^2}+\frac{r}{y}
\frac{\theta'^2}{\sqrt{x'^2+r'^2+y'^2+y^2\theta'^2}}
-\del_{\sigma}\left(\frac{r}{y^2}\frac{y'}{\sqrt{x'^2+r'^2+y'^2+y^2\theta'^2}}\right)=0,\label{eom-y}\\
& \del_{\sigma}\left(\frac{r}{y^2}\frac{x'}{\sqrt{x'^2+r'^2+y'^2+y^2\theta'^2}}\right)=0,\label{eom-x}\\
&\del_{\sigma}\left(\frac{r\theta'}{\sqrt{x'^2+r'^2+y'^2+y^2\theta'^2}}\right)
=0.\label{eom-theta}
\end{align}
Following \cite{Zarembo:1999bu,Olesen:2000ji}, 
we integrate \eqref{eom-x}, \eqref{eom-theta} and fix the reparameterization
of $\sigma$ as $x=\sigma$ to obtain 
\begin{align}
& \frac{r}{y^2}\frac{1}{\sqrt{1+r'^2+y'^2+y^2\theta'^2}}=\ell,\label{ell}\\
& \frac{r\theta'}{\sqrt{1+r'^2+y'^2+y^2\theta'^2}}=m,\label{m}
\end{align}
where $\ell$ and $m$ are integration constants. We can assume $\ell >0$
without loss of generality. Plugging these into \eqref{eom-r} and 
\eqref{eom-y}, we obtain
\begin{align}
 \frac{r}{\ell^2y^4}-r''&=0,
 \label{eom-r2}\\
 -\frac{2r^2}{\ell^2 y^5}+\frac{m^2}{\ell^2 y^3}-y''&=0,
 \label{eom-y2}
\end{align}
and they combine to be 
\begin{align}
 &0=1+r'^2+y'^2-\frac{r^2}{\ell^2y^4}+\frac{m^2}{\ell^2 y^2}.
 \label{read}
\end{align}
Adding \eqref{eom-r2} multiplied by $r$ to 
\eqref{eom-y2} multiplied by $y$ leads to
\begin{align}
& (r^2+y^2)''+2=0.
\end{align}
This equation is integrated to be 
\begin{align}
 &r^2+y^2+(x+c)^2=a^2,
\end{align}where 
$a$ and $c$ are integration constants. 
At $x=0$ (D5-brane side), $r^2+y^2=R_1^2$ leads to 
\begin{align}
 a^2-c^2=R_1^2,
\end{align}
while at $x=L $ (AdS boundary,) $r=R_2$ and $y=0$ lead to
\begin{align}
 a^2-(L +c)^2=R_2^2.
\end{align}
Based on these, $a$ and $c$ can be expressed in terms of $R_1,R_2,L $ as 
\begin{align}
 c=\frac{R_1^2-R_2^2-L ^2}{2L },\qquad
 a=\frac{\sqrt{(R_1^2+R_2^2+L ^2)^2-4R_1^2R_2^2}}{2L }.
\label{ac}
\end{align}
By using the trigonometric parameterization
\begin{align}
 r=\sqrt{a^2-(x+c)^2}\cos\phi ,~~~~
 y=\sqrt{a^2-(x+c)^2}\sin\phi , \qquad
0\le \phi \le \frac{\pi}{2},
 \label{tri}
\end{align}
\eqref{read} becomes 
\begin{align}
\phi'=\pm \left(\frac{a}{a^2-(x+c)^2}\right)
\frac{\sqrt{\cos^2\phi-m^2\sin^2\phi-\ell^2a^2\sin^4\phi}}{\ell a \sin^2\phi}.
 \label{prime}
\end{align}
Furthermore, the boundary condition \eqref{tb} can 
be rewritten as a constraint on 
$\phi_1:=\phi(0)$ and $\phi'(0)$ as
\begin{align}
 \phi'(0)=\cos\theta_k \frac{\sin \phi_1}{\ell R_1^2\cos^2 \phi_1},
\label{phiprime0}
\end{align}
where we have used \eqref{tri}. 
Also, 
\eqref{phiprime0} and \eqref{prime} lead to
the expression of $\phi_1$ as
\begin{align}
 \sin^2 \phi_1=\frac{-(m^2+\sin^2\theta_k)+\sqrt{(m^2+\sin^2\theta_k)^2
 +4\ell^2a^2\sin^2\theta_k}}{2\ell^2 a^2}.\label{sin2phi1}
\end{align}
Finally, \eqref{m} can be rewritten as 
\begin{align}
 \theta'=\frac{m}{\ell(a^2-(x+c)^2)\sin^2\phi}
 \label{thetap}
\end{align}
by using \eqref{ell} and \eqref{tri}. 
This and \eqref{prime} together lead to 
\begin{align}
 \frac{d\phi}{d\theta}=\frac{\phi'}{\theta'}=\pm \frac{1}{m}
 \sqrt{\cos^2\phi-m^2\sin^2\phi-\ell^2 a^2 \sin^4\phi}.
\label{dphidtheta}
\end{align}

Now, we have almost done. Note that \eqref{prime} and \eqref{dphidtheta} can 
be easily integrated. The remaining subtlety is to choose 
the right branch from them. 
At $x=0$, we find that the sign of $\phi'(0)$ is determined 
by the sign of $\cos \theta_k$ from \eqref{phiprime0}. 
We will examine two cases: $\theta_k > \pi/2$ and $\theta_k < \pi/2$ 
in turn. The case of $\theta_k=\pi/2$ is realized as the limit of either case.

\subsection{$\theta_k > \pi/2$: monotonically decreasing $\phi$}
When $\theta_k>\pi/2$, $\phi'(0)$ is
negative by virture of \eqref{phiprime0}, i.e. around $x=0$, $\phi$
decreases. Defining what inside the square root in \eqref{prime}
as a function of $\phi$~: 
\begin{align}
 \Hcal(\phi):=\cos^2\phi-m^2\sin^2\phi-\ell^2 a^2 \sin^4\phi,
\label{h}
\end{align} 
we see that 
$\Hcal(\phi)$ is a monotonically decreasing function on $0 \le \phi \le \pi/2$, 
and $\Hcal(\phi_1)> 0$ according to \eqref{sin2phi1}. 
Hence as $\phi$ decreases, $\Hcal$ becomes bigger and never reaches 
the branching point $\Hcal=0$. 
Therefore, $\phi(x)$ is a monotonically decreasing
function over $0\le x \le L $. We thus take the minus sign
in \eqref{prime}, \eqref{dphidtheta}, and then integrate 
them to obtain 
\begin{align}
& \int_{\phi_1}^{\phi}\frac{\ell a \sin^2\psi d\psi}{\sqrt{\cos^2 \psi-m^2\sin^2\psi-\ell^2 a^2\sin^4 \psi}}
=-\frac12 \log \frac{a+x+c}{a-x-c}+\frac12 \log\frac{a+c}{a-c}, \label{sol1p}\\
&\int_{\phi_1}^{\phi}\frac{m d\psi}{\sqrt{\cos^2 \psi-m^2\sin^2\psi-\ell^2 a^2\sin^4 \psi}}
=-\theta+\theta_k.\label{sol1t}
\end{align}
Note that $\phi(x)$
 and $\theta(x)$ 
can be completely fixed by solving the above equations implemented 
 with the boundary conditions :
\begin{align}
 \phi(L )=0,\qquad \theta(L )=\th,
 \label{bcxi}
\end{align}
which in turn determine $m$ and $\ell$. 
To see this more explicitly, 
it is convenient to introduce 
\begin{align}
f(m,\ell a)&=
\int_0^{\phi_1}\frac
{\ell a \sin^2\psi d\psi}{\sqrt{\cos^2 \psi-m^2\sin^2\psi -\ell^2 a^2\sin^4 \psi}} ,\label{f}\\
g(m,\ell a)&=\int_0^{\phi_1} \frac{m d\psi }{\sqrt{\cos^2 \psi-m^2\sin^2\psi 
-\ell^2 a^2\sin^4 \psi}}.
\label{g}
\end{align}
Rewriting the L.H.S. of \eqref{sol1p} at $x=L $ via 
\eqref{ac} into 
\begin{align}
\begin{aligned}
-\frac{1}{2}\log\frac{a+L +c}{a-L -c}
+\frac{1}{2}\log\frac{a+c}{a-c}
&=-\log\frac{R_1^2+R_2^2+L ^2+\sqrt{(R_1^2 -R_2^2)^2+L ^4+2L ^2(R_1^2 +R_2^2)}}
{2R_1 R_2}\\
&=-\log\left[\eta+\sqrt{\eta^2-1}\right],
\end{aligned}
\end{align}
where $\eta$ is defined in \eqref{eta}, 
we can re-express \eqref{bcxi} as 
\begin{align}
\cosh[f(m,\ell a)]=\eta,\qquad
g(m,\ell a)= \th-\theta_k.
\label{Theta}
\end{align}
Namely, 
$m$ and $\ell a$ are determined in terms of $\eta,\th$ if exist. 
Let us compute the on-shell action. 
The bulk action \eqref{bulk-action} now takes the form 
\begin{align}
&S_{bulk}=\sqrt{\la}\int_0^L  dx \frac{r}{y^2}
 \sqrt{1+r'^2+y'^2 +y^2\theta'^2} 
=\sqrt{\la}\int^{\phi_1}_{\frac{\epsilon}{R_2}}
\frac{d\phi\cot^2\phi}{\sqrt{\cos^2\phi -m^2 \sin^2\phi -{\ell^2 a^2} \sin^4\phi}},
\end{align}
where we have put a cutoff at $y=\epsilon$. We ignore terms which vanish
when $\epsilon \to 0$. Also, there are two boundary terms
\begin{align}
&x=0, ~~~~S_{bdy,1}=\sqrt{\lambda }\cos\theta _k (1-\csc\phi_1),
\label{onbdy1}\\
&x=L, ~~~~S_{bdy,2}=-\sqrt{\lambda}\frac{R_2}{\epsilon},
\label{onbdy2}
\end{align}
from \eqref{gaugevol} and \eqref{bt2}, respectively. 
The total on-shell action is $S_{tot}=S_{bulk}+S_{bdy,1}+S_{bdy,2}$.

Here we summarize the result of this subsection. When $\theta_k > \pi/2$,
the on-shell action can be written by using the formulas in appendix \ref{app-elliptic} as
\begin{align}
\begin{aligned}
\frac{S_{tot}}{\sqrt{\lambda}}=&
[(m^2+1)^2+4\ell^2 a^2]^{1/4}
\Bigg[-\cot\chi(\phi_1)\sqrt{1-\kappa^2\sin^2\chi(\phi_1)}-E(\chi(\phi_1),\kappa)\\
&\qquad \qquad\qquad \qquad
+(1-\kappa^2)F(\chi(\phi_1),\kappa)\Bigg]
+\cos\theta _k (1-\csc\phi_1).
\end{aligned}
\label{res1s}
\end{align}
Note that $\phi_1$ is defined in \eqref{sin2phi1}, whereas 
$\chi(\phi), \kappa$ and $C$ 
are defined in \eqref{defi}. Here 
$F$ and $E$ are the elliptic integrals of the first and second kind, respectively. 
The integration constants $m, \ell a$ are determined by
\begin{align}
\arccosh \eta&=\frac{(m^2+1)+\sqrt{(m^2+1)^2+4\ell^2 a^2}}{
2\ell a \left[(m^2+1)^2+4\ell^2 a^2\right]^{1/4}}
\left[
\Pi(\chi(\phi_1),C,\kappa)-F(\chi(\phi_1),\kappa)
\right],\label{res1f}\\
\th-\theta_k&=m[(m^2+1)^2+4\ell^2 a^2]^{-1/4}
F(\chi(\phi_1),\kappa),\label{res1g}
\end{align}
where $\Pi$ is the elliptic integral of the third kind. 
We found that this result satisfies the conformal 
symmetry condition indicated in \eqref{eta}.

\subsection{$\theta_k < \pi/2$: increasing and decreasing $\phi$}
When $\theta_k<\pi/2$, $\phi'(0)$ is positive by 
virtue of \eqref{phiprime0}. Note that 
$\phi$ keeps increasing from $x=0$ 
(during when $\Hcal$ in \eqref{h} decreases) untill 
the branching point ($\Hcal=0$) is reached at $x=x_0$. This causes 
a sign change in 
\eqref{prime} and \eqref{dphidtheta} such that 
$\phi$ starts to decrease all the way to zero at $x=L $. 
Consequently, we take the plus branch for $x\le x_0$, and the minus branch
for $x \ge x_0$ in \eqref{prime} and \eqref{dphidtheta}.

It is convenient to define $\phi_0:=\phi(x_0)$, 
which satisfies $\Hcal(\phi_0)=0$ and can be solved as
\begin{align}
 \sin^2 \phi_0=\frac{-(m^2+1)+\sqrt{(m^2+1)^2+4\ell^2 a^2}}{2\ell^2 a^2}.
\label{phi0}
\end{align}
When $x\le x_0$, integrating \eqref{prime} and \eqref{dphidtheta} gives
\begin{align}
& \int_{\phi_1}^{\phi}\frac{\ell a \sin^2\psi d\psi}{\sqrt{\cos^2 \psi-m^2\sin^2\psi-\ell^2 a^2\sin^4 \psi}}
=\frac12 \log \frac{a+x+c}{a-x-c}-\frac12 \log\frac{a+c}{a-c}, \label{sol21p}\\
&\int_{\phi_1}^{\phi}\frac{m d\psi}{\sqrt{\cos^2 \psi-m^2\sin^2\psi-\ell^2 a^2\sin^4 \psi}}
=\theta-\theta_k.\label{sol21t}
\end{align}
On the other hand, when $x\ge x_0$, 
\begin{align}
& \left(\int_{\phi_1}^{\phi_0}-\int_{\phi_0}^{\phi}\right)\frac{\ell a \sin^2\psi d\psi}{\sqrt{\cos^2 \psi-m^2\sin^2\psi-\ell^2 a^2\sin^4 \psi}}
=\frac12 \log \frac{a+x+c}{a-x-c}-\frac12 \log\frac{a+c}{a-c}, \label{sol22p}\\
&\left(\int_{\phi_1}^{\phi_0}-\int_{\phi_0}^{\phi}\right)\frac{m d\psi}{\sqrt{\cos^2 \psi-m^2\sin^2\psi-\ell^2 a^2\sin^4 \psi}}
=\theta-\theta_k.\label{sol22t}
\end{align}

We again define 
\begin{align}
\tilde f(m,\ell a)&=\left(\int_{\phi_1}^{\phi_0}+\int_{0}^{\phi_0}\right)
\frac{\ell a \sin^2\psi d\psi}{\sqrt{\cos^2 \psi-m^2\sin^2\psi -{\ell^2 a^2}\sin^4 \psi}} ,\\
\tilde g(m,\ell a)&=\left(\int_{\phi_1}^{\phi_0}+\int_{0}^{\phi_0}\right)
\frac{m d\psi }{\sqrt{\cos^2 \psi-m^2\sin^2\psi -{\ell^2 a^2}\sin^4 \psi}}
\end{align}
as before in order to incorporate the boundary condition \eqref{bcxi}. 
\eqref{sol22p}, \eqref{sol22t} and \eqref{bcxi} lead to
\begin{align}
\eta=\cosh [\tilde f(m,\ell a)],\qquad
\tilde g(m,\ell a)= \th-\theta_k.
\label{Theta--}
\end{align}
Here, $m$ and $\ell a$ are determined in terms of $\eta,\th$ if exist. 
The on-shell action $S_{tot}$ contains 
the bulk part 
\begin{align}
\begin{aligned}
 S_{bulk}
&=\sqrt{\la}\left(\int_{\phi_1}^{\phi_0}
+\int_{\frac{\epsilon}{R_2}}^{\phi_0}\right)
\frac{d\phi\cot^2\phi}{\sqrt{\cos^2\phi -m^2 \sin^2\phi -{\ell^2 a^2} \sin^4\phi}}
\label{Sinde}
\end{aligned}
\end{align}
as well as boundary terms which are the same as \eqref{onbdy1} and \eqref{onbdy2}.

We summarize the result of this subsection. When $\theta_k < \pi/2$,
the on-shell action can be written by using the formulas in appendix \ref{app-elliptic} as
\begin{align}
\begin{aligned}
\frac{S_{tot}}{\sqrt{\lambda}}=&
[(m^2+1)^2+4\ell^2 a^2]^{1/4}
\Bigg[\cot\chi(\phi_1)\sqrt{1-\kappa^2\sin^2\chi(\phi_1)}
\\&
+E(\chi(\phi_1),\kappa)
-(1-\kappa^2)F(\chi(\phi_1),\kappa)-2E(\kappa)
+2(1-\kappa^2)K(\kappa)\Bigg]
+\cos\theta _k (1-\csc\phi_1),
\end{aligned}
\label{res2s}
\end{align}
where $\phi_1$ is defined in \eqref{sin2phi1}, while $\chi(\phi), \kappa$, and 
$C$ are defined in \eqref{defi}. 
The integration constants $m, \ell a$ are determined by
\begin{align}
& \arccosh \eta=\frac{(m^2+1)+\sqrt{(m^2+1)^2+4\ell^2 a^2}}{
2\ell a \left[(m^2+1)^2+4\ell^2 a^2\right]^{1/4}}
\left[
2\Pi(C,\kappa)-2K(\kappa)
-\Pi(\chi(\phi_1),C,\kappa)+F(\chi(\phi_1),\kappa)
\right],\label{res2f}\\
& \th-\theta_k=m[(m^2+1)^2+4\ell^2 a^2]^{-1/4}[
2K(\kappa)
-F(\chi(\phi_1),\kappa)].
\label{res2g}
\end{align}
Again, the conformal symmetry condition in 
\eqref{eta} is satisfied.

\section{Comments on the solutions}
\subsection{Gross-Ooguri phase transition}
Let us consider the Gross-Ooguri (GO) phase transition here. We compare 
\eqref{res1s}-\eqref{res1g} (or \eqref{res2s}-\eqref{res2g}, when 
$\theta_k < \pi/2$) and \eqref{sdisk}. We are also interested in 
how far the annulus solution can stretch along $\eta$. For example, when $\theta_k > \pi/2$, 
the annulus solution exists, if and only if 
\eqref{res1f} and \eqref{res1g} can be solved by some $(m,\ell a)$ 
for a given $(\eta,\th)$ pair.

\begin{figure}[htbp]
\begin{center}
 \includegraphics{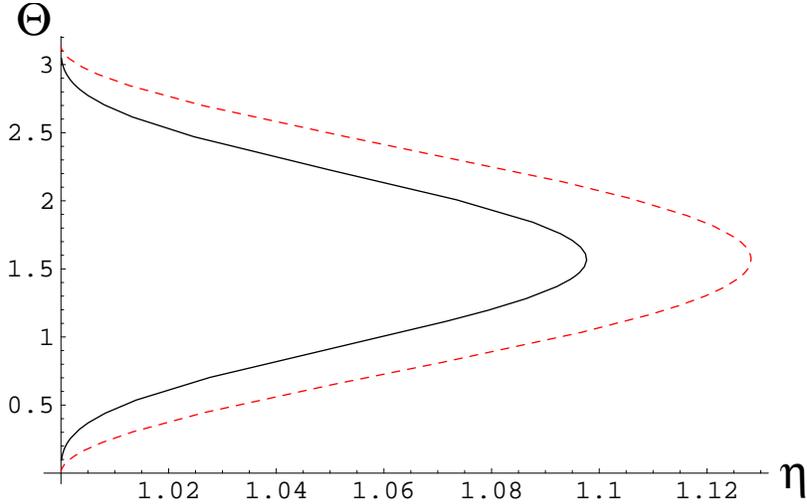}
\end{center}
\caption{Phase diagram on $\eta$-$\th$ plane with $\theta_k=\pi/2$.  The
solid black line indicates the critical line of the GO transition.
Inside (outside) the solid line, the annulus (disk) solution dominates
the connected correlator.  The red dashed line represents the critical
line where the annulus solution becomes unstable. The annulus solution
does not exist outside the red dashed line.}  \label{pi/2}
\end{figure}

\begin{figure}[htbp]
\begin{center}
 \includegraphics{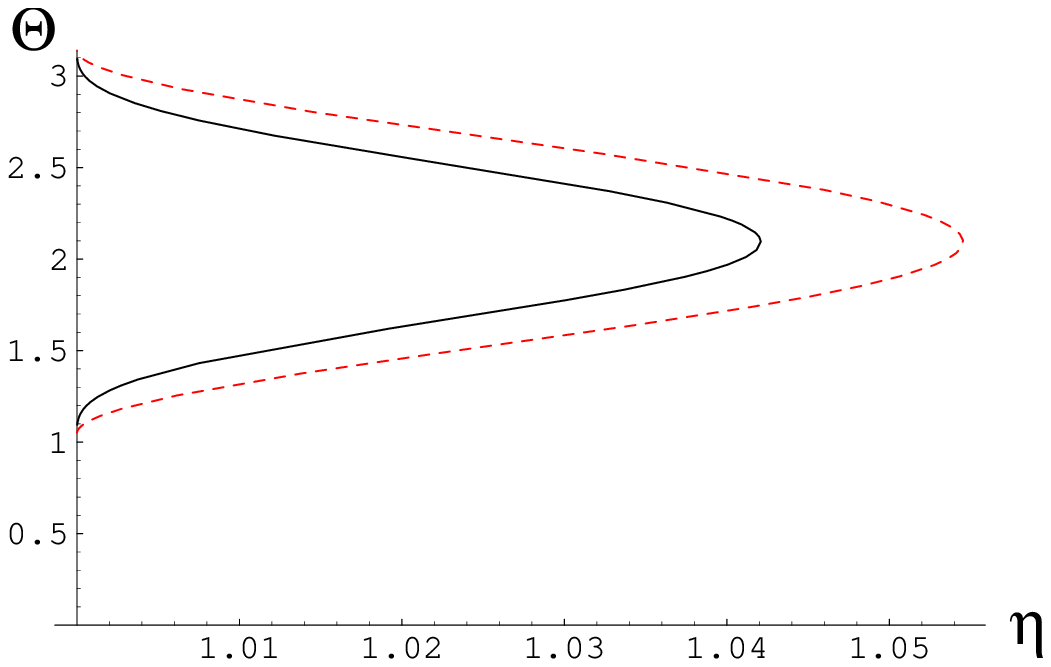}
\end{center}
\caption{Phase diagram on $\eta$-$\th$ plane with $\theta_k=2\pi/3$.}
\label{2pi/3}
\end{figure}

\begin{figure}[htbp]
\begin{center}
 \includegraphics{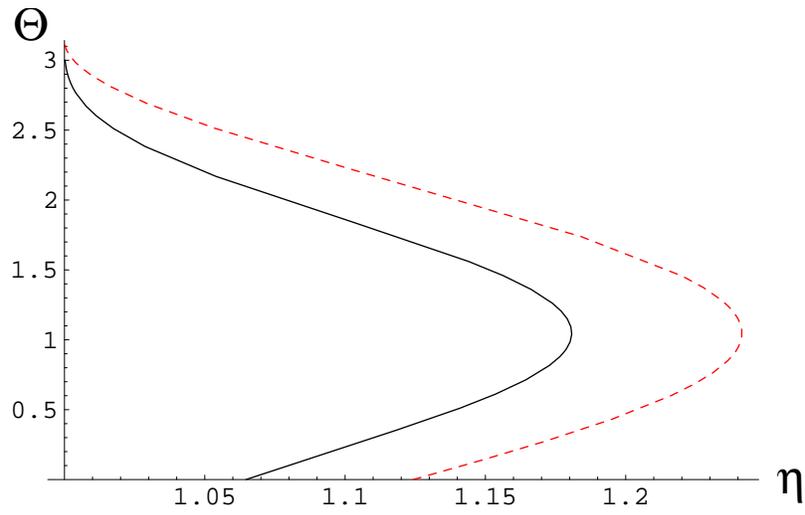}
\end{center}
\caption{Phase diagram on $\eta$-$\th$ plane with $\theta_k=\pi/3$.}
\label{pi/3}
\end{figure}

Figures \ref{pi/2}, \ref{2pi/3} and \ref{pi/3} present the phase
diagrams plotted against the $\eta$-$\th$ plane at $\theta_k=\pi/2$, $\
2\pi/3$ and $\ \pi/3$, respectively. The black solid line indicates the
critical line of the GO phase transition. 
The red dashed line stands for the critical line where the annulus solution 
becomes unstable.

One finds that $\th=\pi$ is always in the disk phase.  This
observation has a good explanation. When $\th=\pi$, two loops 
get overlapped ($\eta=1$) so that this configuration becomes BPS. Therefore, the
annulus solution does not exist.

The solution obtained in the previous section has a ``duality'' for a fixed
$k$, i.e. $(\th,\eta)\to(2\theta_k-\th,\eta)$.
Consequently, for $k\ge N/2$ ($\theta_k\ge \pi/2$), 
$\th\le 2\theta_k-\pi$ is 
always in the disk phase. 
In paticular, the point $\th=2\theta_k-\pi,\ \eta=0$
bears a similar nature to the BPS point $\th=\pi,\ \eta=0$ in the approximation used here.

In addition, $\th$ dependence of the phase diagram is interesting.
The critical distance marks a maximum at $\th=\theta_k$ when $\theta_k$
is fixed.  This fact is quite reasonable from the AdS point of
view. However, it is rather mysterious from the gauge theory side, since 
$\th$ and $\theta_k$ have rather different origins; $\th$ is
the direction of the scalar as in \eqref{c1}, while $\theta_k$ is
determined by the rank of the anti-symmetric representation as
 in \eqref{thetak}.  To consider this fact from the gauge theory is an
interesting future work.

\subsection{$k=1$}
It is illuminating to check some limiting aspects of the above 
solutions. When $k=1$, the $k$-th anti-symmetric representation reduces 
to the fundamental one. Assuming $k=1$, one 
may expect \eqref{res2s}-\eqref{res2g} reproduce the correlator of two 
fundamental loops. We will see this is actually the case. It is 
nontrivial in the sense that the D5-brane picture of the 
anti-symmetric loop is valid only when $k$ is large and comparable to 
$N$. From \eqref{thetak}, it is seen that 
$\lim_{k\rightarrow1}\theta_{k}\cong \left(\frac{3\pi}{2N}\right)^{1/3} 
\ll 1$. Due to $\theta_k\ll 1$, one finds $\phi_1 \ll 1$ by
using \eqref{sin2phi1}. This means that the endpoint on the D5-brane is 
drawn closely to the AdS boundary.

Taking these facts into account, we can express the results
 \eqref{res2s}-\eqref{res2g} for $k=1$ as

\begin{align}
\begin{aligned}
S_{tot}=&
\sqrt{\lambda}+2\sqrt{\lambda}[(m^2+1)^2+4\ell^2 a^2]^{1/4}
\left[
-E(\kappa)
+(1-\kappa^2)K(\kappa)\right]\label{fk1}
\end{aligned}
\end{align}
\begin{align}
& \arccosh \eta=\frac{(m^2+1)+\sqrt{(m^2+1)^2+4\ell^2 a^2}}{
\ell a \left[(m^2+1)^2+4\ell^2 a^2\right]^{1/4}}
\left[
\Pi(C,\kappa)-K(\kappa)
\right],\label{res2f}\\
& \th=2m[(m^2+1)^2+4\ell^2 a^2]^{-1/4}
K(\kappa).
\end{align}

%\begin{align}
% \arccosh \eta=
% \frac{1+\sqrt{1+4\ell^2 a^2}}{\ell a(1+4\ell^2 a^2)^{1/4}}\left[
%\Pi(C,\kappa)-K(\kappa)
%\right], ~~~~
%\kappa^2=\frac12+\frac{1}{2\sqrt{1+4\ell^2 a^2}},\label{fk1}
%\end{align}
%\begin{align}
%S_{tot}=\sqrt{\lambda}+2\sqrt{\lambda} \left[1+4\ell^2 a^2\right]^{1/4}
%\left[-E(\kappa)+(1-\kappa^2)K(\kappa)\right]\label{Sk1}.
%\end{align}

Recall that $S_{tot}$ is identified with $-\log V_k$ in \eqref{logV}. 
The first term in $S_{tot}$ is interpreted as the 
contribution from the denominator in \eqref{Vk}. The second term 
in \eqref{fk1} is 
completely the same as the one in the correlator of 
two fundamental Wilson loops, 
see \cite{Drukker:2005cu}. 
If one takes further the limit $\th\to 0$ 
in the above expression, it reproduces the results of 
\cite{Zarembo:1999bu,Olesen:2000ji, Kim:2001td}.

\subsection{Limit to anti-parallel lines}\label{sec-coulomb}

When $R_1=R_2 \gg L $, the Wilson loop correlator 
reduces to the anti-parallel lines. In this case, one expects that the 
correlator exhibits the Coulomb's law as
\begin{align}
 -\log V_{k}(R_1,R_2,L ,\th)\cong 2\pi R_2 \frac{M}{L },
\end{align}
where $M$ is a constant determined by $k$ and $\th$. 
In terms of $\ell$ and $m$, this limit is 
realized via $\ell a \to \infty$ with $\gamma=\frac{m^2}{2\ell a}$ fixed. 
The explicit form of the coefficient $M$ is written down in appendix
 \ref{app-coulomb}.

%%%%%%%%%%%%%%%%%%%%%%%%%%%%%%%%%%%%%%%%%%%%%%%%%%%%%%%%%%%%%%%%%%%%%
\subsection*{Acknowledgements}
We would like to thank Changhyun Ahn, Yosuke Imamura, Feng-Li Lin, 
Yutaka Matsuo, Nikita Nekrasov, Soo-Jong Rey and 
Jung-Tay Yee for useful discussions and comments. The work of S.Y. was 
supported in part by the European Research Training Network contract 
005104 ``ForcesUniverse.'' 
%%%%%%%%%%%%%%%%%%%%%%%%%%%%%%%%%%%%%%%%%%%%%%%%%%%%%%%%%%%%%%%%%%%%%

\appendix
\section{Elliptic integrals}\label{app-a}

\subsection{Definitions and some formulas of the standard elliptic integrals}
\begin{align}
 &F(\vp,k):=\int_0^{\sin \vp} dz \frac{1}{\sqrt{(1-z^2)(1-k^2z^2)}}
   ,\\
 &E(\vp,k):=\int_0^{\sin \vp} dz \sqrt{\frac{1-k^2 z^2}{1-z^2}}
  ,\\
 &\Pi(\vp,C,k):=\int_0^{\sin\vp}dz 
\frac{1}{(1-C z^2)\sqrt{(1-z^2)(1-k^2 z^2)}}.
\end{align}
Complete elliptic integrals are
\begin{align}
 K(k):=F(\pi/2,k),\qquad
 E(k):=E(\pi/2,k),\qquad
 \Pi(C,k):=\Pi(\pi/2,C,k).
\end{align}
When $C=k^2$, the third elliptic integral is related to the second one as
\begin{align}
 \Pi(\vp,k^2,k)=\frac{1}{1-k^2}E(\vp,k)-\frac{k^2}{1-k^2}\frac{\sin\vp\cos\vp}{\sqrt{1-k^2\sin^2\vp}}.
\end{align}
Let $a,b,c,d$ be four real constants which satisfies $a<b<c<d$. We assume the real variable $u$ satisfies $b\le u \le c$. 
If two variable $u$ and $z$ are related as
\begin{align}
 z^2:=\frac{(c-a)(u-b)}{(c-b)(u-a)},\quad
\text{or} \quad
u=\frac{(c-a)b-(c-b)a z^2}{(c-a)-(c-b)z^2},
\end{align}
there is a 1-form relation 
\begin{align}
&\frac{du}{\sqrt{(u-a)(u-b)(u-c)(u-d)}}
=\frac{2}{\sqrt{(c-a)(d-b)}}\frac{dz}{\sqrt{(1-z^2)(1-k^2z^2)}},\\
&k^2:=\frac{(c-b)(d-a)}{(c-a)(d-b)}.
\end{align}
We obtain the following formula.
\begin{align}
& \int_{b}^{v}\frac{du}{\sqrt{(u-a)(u-b)(u-c)(u-d)}}
=\frac{2}{\sqrt{(c-a)(d-b)}} F(\vp,k),\\
& \sin\vp:=\frac{(c-a)(v-b)}{(c-b)(v-a)}.
\end{align}
\begin{align}
 &\int_{b}^{v}du\sqrt{\frac{(u-b)}{(u-a)(u-c)(u-d)}}
 =\frac{2(b-a)}{\sqrt{(c-a)(d-b)}} \left[\Pi(\vp,C,k)-F(\vp,k)\right],\\
&C=\frac{c-b}{c-a}.
\end{align}
Let $\delta$ be a small number.
\begin{align}
 \int_{b+\delta}^{v}
&\frac{1}{(u-b)\sqrt{(u-a)(u-b)(u-c)(u-d)}}
=\nonumber\\
&\frac{2}{\sqrt{(c-a)(d-b)}}
\Bigg\{
\sqrt{\frac{c-a}{(c-b)(b-a)}}\frac{1}{\sqrt{\delta}}
-\frac{(c-a)}{(c-b)(b-a)} \cot \vp\sqrt{1-k^2\sin^2 \vp}
\nonumber\\&\qquad\qquad
-\frac{(c-a)}{(c-b)(b-a)}E(\vp,k)
+\frac{1}{c-b}F(\vp,k)
\Bigg\}+(\text{terms vanishing when } \delta \to 0).
\end{align}
In order to derive this formula, the relation
\begin{align}
 \frac{1}{z^2\sqrt{(1-z^2)(1-k^2z^2)}}
=\frac{d}{dz}\left[\frac{-1}{z}\sqrt{(1-z^2)(1-k^2z^2)}\right]
+\sqrt{\frac{1-k^2z^2}{1-z^2}}
+\sqrt{(1-z^2)(1-k^2z^2)}
\end{align}
is useful.

\subsection{Expressions of some integrals in terms of the standard elliptic integrals}
\label{app-elliptic}
\begin{align}
&u=\sin^2 \psi,\qquad v=\sin^2 \phi,\qquad
\beta_{\pm}=\frac{-(m^2+1)\pm \sqrt{(m^2+1)^2+4\ell^2 a^2}}{2\ell^2 a^2},
\end{align}
where $\beta_{-}<0\le v\le \beta_{+}<1$, and 
\begin{align}
&\frac{d\psi}{\sqrt{\cos^2\psi-m^2\sin^2\psi-\ell^2 a^2\sin^4\psi}}
=\frac{du}{2\ell a\sqrt{(u-\beta_{-})u(u-\beta_{+})(u-1)}}.
\end{align}
\begin{align}
\kappa:=\sqrt{\frac{\beta_{+}(1-\beta_{-})}{\beta_{+}-\beta_{-}}},\qquad
C:=\frac{\beta_{+}}{\beta_{+}-\beta_{-}},\qquad
\chi(\phi):=\sin^{-1}\sqrt{\frac{(\beta_{+}-\beta_{-})\sin^2\phi}{\beta_{+}
(\sin^2\phi-\beta_{-})}}.\label{defi}
\end{align}
\begin{align}
\begin{aligned}
&\int_{0}^{\phi}d\psi\frac{\ell a \sin^2\psi d\psi}
{\sqrt{\cos^2\psi-m^2\sin^2\psi-\ell^2 a^2\sin^4\psi}}
=\frac12 \int_{0}^{v}du\sqrt{\frac{u}{(u-\beta_{-})(u-\beta_{+})(u-1)}}\\
&=\frac{(m^2+1)+\sqrt{(m^2+1)^2+4\ell^2 a^2}}{
2\ell a \left[(m^2+1)^2+4\ell^2 a^2\right]^{1/4}}
\left[
\Pi(\chi(\phi),C,\kappa)-F(\chi(\phi),\kappa)
\right].
\end{aligned}
\end{align}
\begin{align}
\begin{aligned}
&\int_{0}^{\phi}d\psi\frac{m d\psi}{\sqrt{\cos^2\psi-m^2\sin^2\psi-\ell^2 a^2\sin^4\psi}}
=\frac1{2\ell a} \int_{0}^{v}du\frac{m}{\sqrt{(u-\beta_{-})u(u-\beta_{+})(u-1)}}\\
&=m[(m^2+1)^2+4\ell^2 a^2]^{-1/4}
F(\chi(\phi),\kappa).
\end{aligned}
\end{align}
\begin{align}
\begin{aligned}
& \int_{\frac{\epsilon}{R_2}}^{\phi}d\psi\frac{\cot^2 \psi d\psi}
{\sqrt{\cos^2\psi-m^2\sin^2\psi-\ell^2 a^2\sin^4\psi}}
=\frac1{2\ell a} \int_{\frac{\epsilon^2}{R_2^2}}^{v}du\frac{1-u}{u\sqrt{(u-\beta_{-})u(u-\beta_{+})(u-1)}}\\
&=\frac{R_2}{\epsilon}+[(m^2+1)^2+4\ell^2 a^2]^{1/4}
\left[-\cot\chi(\phi)\sqrt{1-\kappa^2\sin^2\chi(\phi)}-E(\chi(\phi),\kappa)
+(1-\kappa^2)F(\chi(\phi),\kappa)\right].
\end{aligned}
\end{align}
When $\phi=\phi_0$, each integral becomes a complete elliptic integral, i.e. 
\begin{align}
 \int_{0}^{\phi_0}d\psi\frac{\ell a \sin^2\psi d\psi}{\sqrt{\cos^2\psi-m^2\sin^2\psi-\ell^2 a^2\sin^4\psi}}
&=\frac{(m^2+1)+\sqrt{(m^2+1)^2+4\ell^2 a^2}}{
2\ell a \left[(m^2+1)^2+4\ell^2 a^2\right]^{1/4}}
\left[
\Pi(C,\kappa)-K(\kappa)
\right],
\label{Appi0}
\end{align}
\begin{align}
 \int_{0}^{\phi_0}d\psi\frac{m d\psi}{\sqrt{\cos^2\psi-m^2\sin^2\psi-\ell^2 a^2\sin^4\psi}}
&=m[(m^2+1)^2+4\ell^2 a^2]^{-1/4}
K(\kappa),
\label{Appi1}
\end{align}
\begin{align}
& \int_{\frac{\epsilon}{R_2}}^{\phi_0}d\psi\frac{\cot^2 \psi d\psi}
{\sqrt{\cos^2\psi-m^2\sin^2\psi-\ell^2 a^2\sin^4\psi}}
&=\frac{R_2}{\epsilon}+[(m^2+1)^2+4\ell^2 a^2]^{1/4}
\left[-E(\kappa)
+(1-\kappa^2)K(\kappa)\right].
\label{Appi2}
\end{align}

\section{Explicit form of the coulomb coefficient}
\label{app-coulomb}
Here we write the explicit form mentioned in section \ref{sec-coulomb}.

\begin{align}
\begin{aligned}
 \beta_{\pm}&=\frac{-\gamma \pm \sqrt{\gamma^2+1}}{\ell a},\qquad 
\kappa^2=C=\frac{-\gamma+\sqrt{\gamma^2+1}}{2\sqrt{\gamma^2+1}},\qquad
\sin^2\phi_1=\frac{-\gamma + \sqrt{\gamma^2+\sin^2\theta_k}}{\ell a},\\
\chi(\phi_1)&=\sin^{-1}\sqrt{
\frac{2(-\gamma+\sqrt{\gamma^2+\sin^2\theta_k})\sqrt{\gamma^2+1}}{
(-\gamma+\sqrt{\gamma^2+1})(\sqrt{\gamma^2+\sin^2\theta_k}+\sqrt{\gamma^2+1})}}
,\qquad
 \eta=1+\frac12 \left(\frac{L }{R_2}\right)^2.      
\end{aligned}
\label{anti}
\end{align}

\subsection{$\theta_k >\pi/2$}
$\gamma$ is related to $\th$ by
\begin{align}
\th-\theta_k=\frac{\sqrt{\gamma}}{(\gamma^2+1)^{1/4}}
 F(\chi(\phi_1),\kappa).
\end{align}
The potential is
\begin{align}
 S_{tot}=2\pi R_2 \frac{M}{L },
\end{align}
where the coefficient can be written as
\begin{align}
 M=&\frac{1}{2\pi}\frac{\gamma+\sqrt{\gamma^2+1}}{\sqrt{2}(\gamma^2+1)^{1/4}}
\left[\Pi(\chi(\phi_1),\kappa^2,\kappa)-F(\chi(\phi_1),\kappa)\right]\nonumber\\
&\times\sqrt{\lambda}\Bigg[
-\frac{\cos \theta_k}{\sqrt{-\gamma+\sqrt{\gamma^2+\sin^2\theta_k}}}
\nonumber\\
&
+\sqrt{2}(\gamma^2+1)^{1/4}\left\{
-\cot\chi(\phi_1)\sqrt{1-\kappa^2\sin^2\chi(\phi_1)}
-E(\chi(\phi_1),\kappa)+(1-\kappa^2)F(\chi(\phi_1),\kappa)
\right\}\Bigg].
\end{align}

\subsection{$\theta_k<\pi/2$}

\begin{align}
&\th-\theta_k=\frac{\sqrt{\gamma}}{(\gamma^2+1)^{1/4}}
\left[2K(\kappa)-F(\chi(\phi_1),\kappa)\right] .
\end{align}

\begin{align}
 M=&\frac{1}{2\pi}\frac{\gamma+\sqrt{\gamma^2+1}}{\sqrt{2}(\gamma^2+1)^{1/4}}
\left[
2\Pi(\kappa^2,\kappa)-\Pi(\chi(\phi_1),\kappa^2,\kappa)
-2K(\kappa)+F(\chi(\phi_1),\kappa)
\right]\nonumber\\
&\times\sqrt{\lambda}\Bigg[
-\frac{\cos \theta_k}{\sqrt{-\gamma+\sqrt{\gamma^2+\sin^2\theta_k}}}
+\sqrt{2}(\gamma^2+1)^{1/4}\Big\{
\cot\chi(\phi_1)\sqrt{1-\kappa^2\sin^2\chi(\phi_1)}
\nonumber\\
&
-2E(\kappa)
+E(\chi(\phi_1),\kappa)
+2(1-\kappa^2)K(\kappa)
-(1-\kappa^2)F(\chi(\phi_1),\kappa)
\Big\}\Bigg].
\end{align}

%%%%%%%%%%%%%%%%%%%%%%%%%%%%%%%%%%%%%%%%%%%%%%%%%%%%%%%%%%%%%%%%


\begin{thebibliography}{10}

\bibitem{Maldacena:1997re}
J.~M. Maldacena, {\it ``The large {N} limit of superconformal field theories
  and supergravity,''}  Adv. Theor. Math. Phys. {\bf 2} (1998) 231--252,
  [\href{http://arXiv.org/abs/hep-th/9711200}{{\tt hep-th/9711200}}].

\bibitem{Rey:1998ik}
S.-J. Rey and J.-T. Yee, {\it ``Macroscopic strings as heavy quarks in large
  {N} gauge theory and anti-de {S}itter supergravity,''}  Eur. Phys. J. {\bf
  C22} (2001) 379--394, [\href{http://arXiv.org/abs/hep-th/9803001}{{\tt
  hep-th/9803001}}].

\bibitem{Maldacena:1998im}
J.~M. Maldacena, {\it ``{W}ilson loops in large {N} field theories,''}  Phys.
  Rev. Lett. {\bf 80} (1998) 4859--4862,
  [\href{http://arXiv.org/abs/hep-th/9803002}{{\tt hep-th/9803002}}].

\bibitem{Drukker:2005kx}
N.~Drukker and B.~Fiol, {\it ``All-genus calculation of {W}ilson loops using
  {D}-branes,''}  JHEP {\bf 02} (2005) 010,
  [\href{http://arXiv.org/abs/hep-th/0501109}{{\tt hep-th/0501109}}].

\bibitem{Hartnoll:2006hr}
S.~A. Hartnoll and S.~Prem~Kumar, {\it ``Multiply wound {P}olyakov loops at
  strong coupling,''}  Phys. Rev. {\bf D74} (2006) 026001,
  [\href{http://arXiv.org/abs/hep-th/0603190}{{\tt hep-th/0603190}}].

\bibitem{Yamaguchi:2006tq}
S.~Yamaguchi, {\it ``{W}ilson loops of anti-symmetric representation and
  {D5}-branes,''}  JHEP {\bf 05} (2006) 037,
  [\href{http://arXiv.org/abs/hep-th/0603208}{{\tt hep-th/0603208}}].

\bibitem{Gomis:2006sb}
J.~Gomis and F.~Passerini, {\it ``Holographic {W}ilson loops,''}  JHEP {\bf 08}
  (2006) 074, [\href{http://arXiv.org/abs/hep-th/0604007}{{\tt
  hep-th/0604007}}].

\bibitem{Erickson:2000af}
J.~K. Erickson, G.~W. Semenoff and K.~Zarembo, {\it ``Wilson loops in {N = 4}
  supersymmetric {Y}ang-{M}ills theory,''}  Nucl. Phys. {\bf B582} (2000)
  155--175, [\href{http://arXiv.org/abs/hep-th/0003055}{{\tt hep-th/0003055}}].

\bibitem{Drukker:2000rr}
N.~Drukker and D.~J. Gross, {\it ``An exact prediction of {N = 4} {SUSYM}
  theory for string theory,''}  J. Math. Phys. {\bf 42} (2001) 2896--2914,
  [\href{http://arXiv.org/abs/hep-th/0010274}{{\tt hep-th/0010274}}].

\bibitem{Rodriguez-Gomez:2006zz}
D.~Rodriguez-Gomez, {\it ``Computing {W}ilson lines with dielectric branes,''}
  Nucl. Phys. {\bf B752} (2006) 316--326,
  [\href{http://arXiv.org/abs/hep-th/0604031}{{\tt hep-th/0604031}}].

\bibitem{Okuyama:2006jc}
K.~Okuyama and G.~W. Semenoff, {\it ``Wilson loops in {N} = 4 {SYM} and fermion
  droplets,''}  JHEP {\bf 06} (2006) 057,
  [\href{http://arXiv.org/abs/hep-th/0604209}{{\tt hep-th/0604209}}].

\bibitem{Hartnoll:2006is}
S.~A. Hartnoll and S.~P. Kumar, {\it ``Higher rank {W}ilson loops from a matrix
  model,''}  JHEP {\bf 08} (2006) 026,
  [\href{http://arXiv.org/abs/hep-th/0605027}{{\tt hep-th/0605027}}].

\bibitem{Hartnoll:2006ib}
S.~A. Hartnoll, {\it ``Two universal results for {W}ilson loops at strong
  coupling,''}  \href{http://arXiv.org/abs/hep-th/0606178}{{\tt
  hep-th/0606178}}.

\bibitem{Chen:2006iu}
B.~Chen and W.~He, {\it ``On 1/2-{BPS} {W}ilson-'t {H}ooft loops,''}
  \href{http://arXiv.org/abs/hep-th/0607024}{{\tt hep-th/0607024}}.

\bibitem{Giombi:2006de}
S.~Giombi, R.~Ricci and D.~Trancanelli, {\it ``Operator product expansion of
  higher rank {W}ilson loops from {D}-branes and matrix models,''}
  \href{http://arXiv.org/abs/hep-th/0608077}{{\tt hep-th/0608077}}.

\bibitem{Gross:1998gk}
D.~J. Gross and H.~Ooguri, {\it ``Aspects of large {N} gauge theory dynamics as
  seen by string theory,''}  Phys. Rev. {\bf D58} (1998) 106002,
  [\href{http://arXiv.org/abs/hep-th/9805129}{{\tt hep-th/9805129}}].

\bibitem{Zarembo:1999bu}
K.~Zarembo, {\it ``Wilson loop correlator in the {AdS/CFT} correspondence,''}
  Phys. Lett. {\bf B459} (1999) 527--534,
  [\href{http://arXiv.org/abs/hep-th/9904149}{{\tt hep-th/9904149}}].

\bibitem{Olesen:2000ji}
P.~Olesen and K.~Zarembo, {\it ``Phase transition in {W}ilson loop correlator
  from {AdS/CFT} correspondence,''}
  \href{http://arXiv.org/abs/hep-th/0009210}{{\tt hep-th/0009210}}.

\bibitem{Kim:2001td}
H.~Kim, D.~K. Park, S.~Tamarian and H.~J.~W. Muller-Kirsten, {\it
  ``{G}ross-{O}oguri phase transition at zero and finite temperature: Two
  circular {W}ilson loop case,''}  JHEP {\bf 03} (2001) 003,
  [\href{http://arXiv.org/abs/hep-th/0101235}{{\tt hep-th/0101235}}].

\bibitem{Arutyunov:2001hs}
G.~Arutyunov, J.~Plefka and M.~Staudacher, {\it ``Limiting geometries of two
  circular {M}aldacena-{W}ilson loop operators,''}  JHEP {\bf 12} (2001) 014,
  [\href{http://arXiv.org/abs/hep-th/0111290}{{\tt hep-th/0111290}}].

\bibitem{Drukker:2005cu}
N.~Drukker and B.~Fiol, {\it ``On the integrability of {W}ilson loops in
  {$AdS_5\times S^5$}: Some periodic ansatze,''}  JHEP {\bf 01} (2006) 056,
  [\href{http://arXiv.org/abs/hep-th/0506058}{{\tt hep-th/0506058}}].

\bibitem{Tsuji:2006zn}
A.~Tsuji, {\it ``Holography of {W}ilson loop correlator and spinning
  strings,''}  \href{http://arXiv.org/abs/hep-th/0606030}{{\tt
  hep-th/0606030}}.

\bibitem{Ahn:2006px}
C.~Ahn, {\it ``Two circular {W}ilson loops and marginal deformations,''}
  \href{http://arXiv.org/abs/hep-th/0606073}{{\tt hep-th/0606073}}.

\bibitem{Zarembo:2001jp}
K.~Zarembo, {\it ``String breaking from ladder diagrams in {SYM} theory,''}
  JHEP {\bf 03} (2001) 042, [\href{http://arXiv.org/abs/hep-th/0103058}{{\tt
  hep-th/0103058}}].

\bibitem{Plefka:2001bu}
J.~Plefka and M.~Staudacher, {\it ``Two loops to two loops in {N} = 4
  supersymmetric {Y}ang-{M}ills theory,''}  JHEP {\bf 09} (2001) 031,
  [\href{http://arXiv.org/abs/hep-th/0108182}{{\tt hep-th/0108182}}].

\bibitem{Yamaguchi:2006te}
S.~Yamaguchi, {\it ``Bubbling geometries for half {BPS} {W}ilson lines,''}
  \href{http://arXiv.org/abs/hep-th/0601089}{{\tt hep-th/0601089}}.

\bibitem{Lunin:2006xr}
O.~Lunin, {\it ``On gravitational description of {W}ilson lines,''}  JHEP {\bf
  06} (2006) 026, [\href{http://arXiv.org/abs/hep-th/0604133}{{\tt
  hep-th/0604133}}].

\bibitem{Berenstein:1998ij}
D.~Berenstein, R.~Corrado, W.~Fischler and J.~M. Maldacena, {\it ``The operator
  product expansion for {W}ilson loops and surfaces in the large {N} limit,''}
  Phys. Rev. {\bf D59} (1999) 105023,
  [\href{http://arXiv.org/abs/hep-th/9809188}{{\tt hep-th/9809188}}].

\bibitem{Pawelczyk:2000hy}
J.~Pawelczyk and S.-J. Rey, {\it ``{R}amond-{R}amond flux stabilization of
  {D}-branes,''}  Phys. Lett. {\bf B493} (2000) 395--401,
  [\href{http://arXiv.org/abs/hep-th/0007154}{{\tt hep-th/0007154}}].

\bibitem{Camino:2001at}
J.~M. Camino, A.~Paredes and A.~V. Ramallo, {\it ``Stable wrapped branes,''}
  JHEP {\bf 05} (2001) 011, [\href{http://arXiv.org/abs/hep-th/0104082}{{\tt
  hep-th/0104082}}].

\bibitem{Drukker:1999zq}
N.~Drukker, D.~J. Gross and H.~Ooguri, {\it ``{W}ilson loops and minimal
  surfaces,''}  Phys. Rev. {\bf D60} (1999) 125006,
  [\href{http://arXiv.org/abs/hep-th/9904191}{{\tt hep-th/9904191}}].

\end{thebibliography}
\end{document}